\documentclass[iop]{emulateapj}

\usepackage{natbib}
\usepackage{ulem}
\usepackage{comment}
\usepackage{txfonts}

\usepackage{color} 
    
\usepackage[T1]{fontenc}

\usepackage{bm}

\newcommand{\eg}{e.g., }
\newcommand{\ie}{i.e., }
\newcommand{\Msun}{\ensuremath{M_{\odot}}~}
\newcommand{\Rsun}{\ensuremath{R_{\odot}}~}
\newcommand{\kms}{km~s$^{-1}$ }

\newcommand{\ergs}{erg~s$^{-1}$ }
\newcommand{\years}{yr$^{-1}$ }
\newcommand{\magdays}{mag~day$^{-1}$}

\newcommand{\Nifs}{$^{56}$Ni~}
\newcommand{\MNifs}{M($^{56}$Ni)}

\def\gsim{\mathrel{\rlap{\lower 4pt \hbox{\hskip 1pt $\sim$}}\raise 1pt
\hbox {$>$}}}
\def\lsim{\mathrel{\rlap{\lower 4pt \hbox{\hskip 1pt $\sim$}}\raise 1pt
\hbox {$<$}}}

\newcommand{\si}{{\sc i}~}
\newcommand{\sib}{{\sc i}}
\newcommand{\sii}{{\sc ii}~}

\newcommand{\caii}{[Ca~{\sc ii}]~}
\newcommand{\oi}{[O~{\sc i}]~}
\newcommand{\cao}{\caii/\oi~}

\newcommand{\ej}{{\rm ej}}

\newcommand{\ek}{E_{\rm k}}

\newcommand{\Z}{{\rm 03Z}}

\newcommand{\ebv}{$E$($B-V$)~}

\newcommand{\bvri}{{\it BVRI}-band }
\newcommand{\jhk}{{\it JHK$_{s}$}-band }

\shorttitle{SN~2019\MakeLowercase{ehk}}
\shortauthors{Nakaoka T., et al.}

\begin{document}

\title{Calcium-rich Transient SN~2019ehk in A Star-Forming Environment: Yet Another Candidate for An Ultra-Stripped Envelope Supernova}
\author{
  Tatsuya Nakaoka\altaffilmark{1,2},
  Keiichi Maeda\altaffilmark{3}, 
  Masayuki Yamanaka\altaffilmark{4},
  Masaomi Tanaka\altaffilmark{5},
  Miho Kawabata\altaffilmark{4}, 
  Takashi J. Moriya\altaffilmark{6,7},
  Koji S. Kawabata\altaffilmark{1,2}, 
  Nozomu Tominaga\altaffilmark{8},
  Kengo Takagi\altaffilmark{2},
  Fumiya Imazato\altaffilmark{2},
  Tomoki Morokuma\altaffilmark{9},
  Shigeyuki Sako\altaffilmark{9},
  Ryou Ohsawa\altaffilmark{9},
  Takashi Nagao\altaffilmark{10},
  Ji-an Jiang\altaffilmark{11},
  Umut Burgaz\altaffilmark{3,12}
  Kenta Taguchi\altaffilmark{3},
  Makoto Uemura\altaffilmark{1,2},
  Hiroshi Akitaya\altaffilmark{1,2},
  Mahito Sasada\altaffilmark{1,2},
  Keisuke Isogai\altaffilmark{4},
  Masaaki Otsuka\altaffilmark{4},
  Hiroyuki Maehara\altaffilmark{4},
}

\altaffiltext{1}{Hiroshima Astrophysical Science Center, Hiroshima University, Kagamiyama 1-3-1, Higashi-Hiroshima ,Hiroshima 739-8526, Japan;nakaokat@hiroshima-u.ac.jp}
\altaffiltext{2}{Department of Physical Science, Hiroshima University, Kagamiyama 1-3-1, Higashi-Hiroshima 739-8526, Japan}
\altaffiltext{3}{Department of Astronomy, Kyoto University,Kitashirakawa-Oiwake-cho, Sakyo-ku, Kyoto 606-8502, Japan}
\altaffiltext{4}{Okayama Observatory, Kyoto University, 3037-5 Honjo, Kamogatacho, Asakuchi, Okayama 719-0232, Japan}
\altaffiltext{5}{Astronomical Institute, Graduate School of Science, Tohoku University, Aramaki, Aoba, Sendai 980-8578, Japan}
\altaffiltext{6}{National Astronomical Observatory of Japan, 2-21-1 Osawa, Mitaka, Tokyo 181-8588, Japan}
\altaffiltext{7}{School of Physics and Astronomy, Faculty of Science, Monash University, Clayton, Victoria 3800, Australia}
\altaffiltext{8}{Department of Physics, Faculty of Science and Engineering, Konan University, Kobe, Hyogo 658-8501, Japan}
\altaffiltext{9}{Institute of Astronomy, Graduate School of Science, The University of Tokyo, 2-21-1 Osawa, Mitaka, Tokyo 181-0015, Japan}
\altaffiltext{10}{European Southern Observatory, Karl-Schwarzschild-Str 2, D-85748 Garching b. M¨unchen, Germany}
\altaffiltext{11}{Kavli Institute for the Physics and Mathematics of the Universe (WPI), The University of Tokyo Institutes for Advanced Study, The University of Tokyo, 5-1-5 Kashiwanoha, Kashiwa, Chiba 277-8583, Japan}
\altaffiltext{12}{Department of Astronomy and Space Sciences, Ege University, 35100, \.{I}zmir, Turkey}

\begin{abstract}
We present optical and near-infrared observations of SN~Ib~2019ehk.
We show that it evolved to a Ca-rich transient according to its spectral properties and evolution in late phases.
It, however, shows a few distinguishable properties from the canonical Ca-rich transients: a  short-duration first peak in the light curve, high peak luminosity, and association with a star-forming environment.
Indeed, some of these features are shared with iPTF14gqr and iPTF16hgs, which are candidates for a special class of core-collapse SNe (CCSNe): the so-called ultra-stripped envelope SNe, \ie a relatively low-mass He (or C+O) star explosion in a binary as a precursor of double neutron star binaries.
The estimated ejecta mass ($0.43 M_\odot$) and explosion energy ($1.7 \times 10^{50} $~erg)
are consistent with this scenario.
The analysis of the first peak suggests existence of dense circumstellar material in the vicinity of the progenitor, implying a CCSN origin.
Based on these analyses, we suggest SN 2019ehk is another candidate for an ultra-stripped envelope SN.
These ultra-stripped envelope SN candidates seem to form a subpopulation among Ca-rich transients, associated with young population. We propose that the key to distinguishing this population is the early first peak in their light curves.

\end{abstract}


\section{Introduction}
\label{sec:intro}










Over the last decade, some peculiar transients
which show different characteristics from canonical supernovae (SNe) have been discovered \citep{kasliwal2012}.
The `Ca-rich' transient is one of such newly discovered types of explosive transients.
Their spectra commonly (if not always) show helium absorption lines around the maximum light and most of these transients are classified into Type Ib SNe (SNe Ib) according to the classical scheme.
However, they gradually start to show different observational properties from the classical SNe Ib after the luminosity maximum  \citep[\eg][]{lunnan2017};
the Ca emission lines quickly develop as early as $\sim$1 month after the maximum, while the oxygen forbidden lines are quite weak.

The origin of the Ca-rich transients is still in active debate.
They are generally discovered in a remote location from a putative host galaxy \citep{kasliwal2012,lunnan2017}, indicating that they originate in old population. 
The old stellar population environment leads to models of white dwarf (WD) explosions \citep{lyman2013,lyman2014_Carich,lyman2016_Carich}, exemplified by the .Ia explosion, \ie helium detonation on a surface of a WD \citep{shen2010}. 

At the discovery of the Ca-rich transient class before this constraint from the environment had been established, one of the first suggestions on their origin was an explosion of a relatively low-mass He or C+O star which represents an SN from the lowest mass range to become core-collapse SNe (CCSNe) \citep{kawabata2010}.
\citet{kawabata2010} pointed out that the dominance of the Ca emissions over the oxygen emissions (\ie deficiency in the oxygen emissions) in their spectra could be understood in the context of an explosion of such a relatively low-mass He (or C+O) progenitor \citep[see][]{maeda2007,fang2019}.
They also argued that the rapid evolution seen in the Ca-rich transient is a natural consequence from such a scenario; this is around a boundary between the SN explosion (either by a Fe-core collapse or ONeMg-core electron capture) and WD formation, and thus predicts small ejecta mass, \ie $< 0.5 M_\odot$ \citep{tauris2013,moriya2017_us}.
The explosion energy and the ejected mass of the newly-formed $^{56}$Ni are also predicted to be small \citep{suwa2015,yoshida2017,muller2018}.
These features are consistent with the observational properties of the Ca-rich transients,
which show low expansion velocities and low luminosity.
In the stellar evolution theory,
one way to from such a low-mass He or C+O progenitor star is a close binary evolution with a neutron star (NS) companion,
which is called the `ultra-stripped envelope SN' (USSN) scenario \citep{tauris2013,tauris2015}.
The USSN scenario is a leading model toward the formation of double NS binaries, and thus identification of USSNe is an important topic in the gravitational wave astronomy as well. 

Given the old stellar environment of the sites where Ca-rich transients are discovered, it is unlikely that the USSN is the origin of a bulk of Ca-rich transients since the USSN should represent a young population.
However, it is still possible that a fraction of Ca-rich transients may come from USSNe. Among some candidates for USSNe \citep{tauris2015,moriya2017_us}, two objects are accompanied with intensive spectral series and multi-band light curves. These two USSN candidates, iPTF16hgs and iPTF14gqr, are indeed either a Ca-rich transient or a peculiar transient which shares some properties with the Ca-rich transient class.
iPTF16hgs was located at $\sim 6$ kpc away from the core of its star-forming host galaxy,
and thus its progenitor can be much younger than most of Ca-rich transients.  \citet{de2018} argued for an USSN origin for iPTF16hgs, while the possibility of a WD eruptive event was not rejected.
Further, \citet{de2018_14ft} suggested that iPTF14gqr,
with some similarities to other Ca-rich transients (Section \ref{sec:result} for details),
is a robust candidate for an USSN.
It was located in a tidally interaction spiral galaxy at $\sim$15~kpc away from the spiral arm (See Section \ref{sec:pop}).




Interestingly, both of iPTF16hgs and iPTF14gqr show double-peaked light curves,
where the first component rapidly declines in a few days to a week. This feature is reminiscent of the so-called `cooling-envelope emission' frequently observed for CCSNe either with an extended envelope or a dense circumstellar material (CSM).
For iPTF16hgs, \citet{de2018} interpreted this as the cooling-envelope emission of a progenitor with a radius of $\sim 10~\Rsun$, which is consistent with a CCSN from a He star and its early spectra classified as an SN Ib.
For iPTF14gqr, \citet{de2018_14ft} suggested that the early emission is associated with a dense and `confined' CSM, as is similar to those frequently inferred for CCSNe \citep{galyam2014,yaron2017,forster2018}.

\begin{figure}[t]
\centering
\includegraphics[width=6cm,clip]{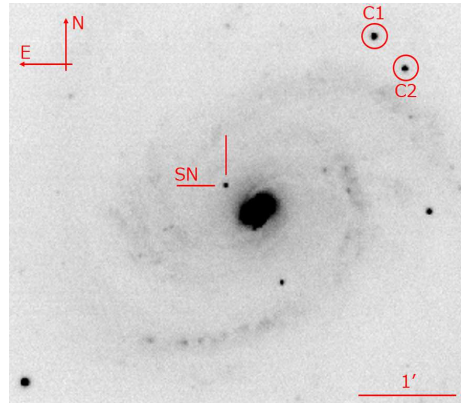}
\caption{The $R$-band image of SN 2019ehk in M~100 (NGC~4321) taken at 
$t = 12.2$ days using HOWPol. The open circles denote the comparison stars 
used for the photometric measurement. The magnitudes of these comparison stars are given by \citet{wang2008}.
}
\label{fig:fc}
\end{figure}

Given the possible link between a fraction of Ca-rich transients and the USSN scenario, even a single new addition of an USSN candidate in the Ca-rich transient class is highly important in both searching for candidate USSNe and in understanding the nature(s) of Ca-rich transients. We here present such a new example, SN 2019ehk. It was discovered at $\alpha$ (J2000)$=12^{\rm h}22^{\rm m}56^{\rm s}.150$,  
$\delta$ (J2000)$=+15^\circ49'34''.03$, in the well-known spiral galaxy M~100 (NGC4321), whose distance is well established through the Cepheid \citep[$m-M = 30.91 \pm 0.14$; ][]{freedman2001}. The projected position of the explosion site is close to the core of NGC4321 (Figure~\ref{fig:fc}). The explosion site is on a dust lane of a spiral arm. The apparently young environment indicates that it is originated from a massive star. Indeed, it has been classified as an SN Ib from early spectroscopic observation \citep{dimitriadis2019}.

In this paper, we present properties of SN 2019ehk. In Section 2, we describe our observations and data reduction.
In Section 3, we present its spectral and light curve properties, and classify it as a Ca-rich transient. We further show that it has a double-peaked light curve. Comparisons of the observational properties of SN~2019ehk and those of iPTF16hgs, iPTF14gqr, and other (canonical) Ca-rich transients are also presented in Section 3. In Section 4, we discuss properties and the origin of SN 2019ehk. We suggest that this is yet another candidate for an USSN, and a fraction of Ca-rich transients belong to this class. The paper is closed in Section 5 with conclusions. 
Throughout this paper, $t$ denotes the rest-frame phase since the $R$-band second maximum. The explosion date is estimated as MJD 58601.9 ($t=-14.4$~days), defined as the date between the last non-detection (MJD 58601.3) and the earliest detection (MJD 58602.5; Joel Shepherd)\footnote{https://wis-tns.weizmann.ac.il/object/2019ehk}.

\section{Observations and Data Reduction}
\label{sec:obs}

\begin{figure}[t]
\centering
\includegraphics[width=8cm,clip]{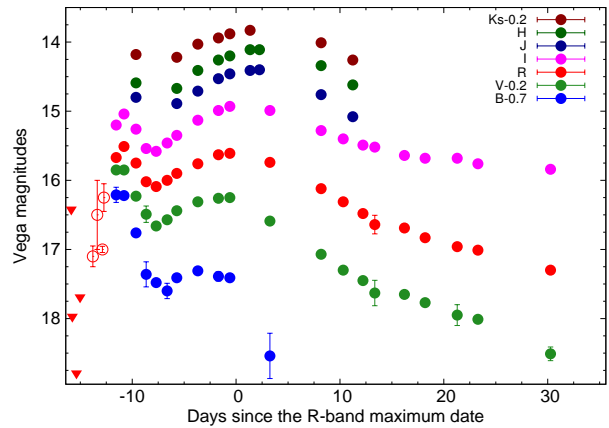}
\caption{The otical and NIR light curves of SN~2019ehk. The estimated explosion date is $t=-14.4$~days (see Section~\ref{sec:intro}). Only the Galactic extinction \citep{schlafly2011} has been corrected for. The host extinction has not been corrected for in this figure (see Section~\ref{sec:extinction}).
}
\label{fig:lc}
\end{figure}

The optical imaging data were obtained using the Hiroshima One-shot Wide-field Polarimeter \citep[HOWPol;][]{kawabata2008} and the Hiroshima Optical and Near-InfraRed Camera \citep[HONIR;][]{akitaya2014}. These instruments are installed on the 1.5-m Kanata telescope at the Higashi-Hiroshima Observatory, Hiroshima University. We obtained \bvri images using HOWPol in 14 nights from 1.7 May 2019 ($t=-11.6$~days) to 12.3 Jul 2019 ($t=30.2$~days). For the photometric measurements, we adopted the Point Spread Function (PSF) photometry task in the {\it DAOPHOT} package \citep{stetson1987} equipped with the {\it IRAF} \citep{tody1986, tody1993}. For the calibration of optical photometry, we used the magnitudes of the nearby comparison stars given by \citet{wang2008}. The derived optical magnitudes are summarized in Table \ref{table:opt}. Figure \ref{fig:lc} shows the multi-band light curves.
Only the Galactic extinction has been corrected for in this figure (see Section~\ref{sec:extinction}). We also plot data from other sources by the open circles (clear-band magnitudes reported by Joel Shepherd and by Jaroslaw Grzegorzek; a cyan-ATLAS-band magnitude by ATLAS)\footnote{https://wis-tns.weizmann.ac.il/object/2019ehk}.

Non-filter optical imaging data in the field around the discovery date 
were taken with a new wide-field CMOS sensor camera Tomo-e Gozen \citep{sako2018}
on the 1.05-m Kiso Schmidt telescope during a wide-field high-cadence transient survey. 
We picked up the survey data on two epochs in the same night of 27 April 2019. 
For each epoch, the data consist of twelve contiguous 0.5 sec exposures. 
SN 2019ehk is not detected, and we derived $5\sigma$ upper limits on the subtracted images, 
using the deep co-added Tomo-e Gozen image taken in 2020 as a reference image. 
The derived upper limits (relative to Pan-STARRS $r$-band) are 16.42 and 17.97 mag on MJD 58600.4 and 58600.5 days, respectively. 

All the magnitudes are given in the Vega system throughout this paper, unless mentioned otherwise.
The double peaks are seen in the optical light curves, and the second peak is reached in the $R$ band at 14.4 days after the estimated explosion date; it is defined as $t=0$~days in this paper. 

The \jhk imaging data were obtained using HONIR in 7 nights since 5.4 May 2019 ($t=-9.7$~days) to 24.5 May 2019 ($t=11.2$~days). We took the images using a dithering mode to accurately subtract a bright foreground sky.
We reduced the data and performed photometry following the standard procedure for NIR data based on 
the PSF photometry method in {\it IRAF}. The magnitudes were calibrated using the magnitudes of nearby comparison stars given in the 2MASS catalog \citep{persson1998}. The derived \jhk magnitudes and their light curves are given in Table \ref{table:nir} and Figure \ref{fig:lc}, respectively.

\begin{figure}[t]
\centering
\includegraphics[width=8cm,clip]{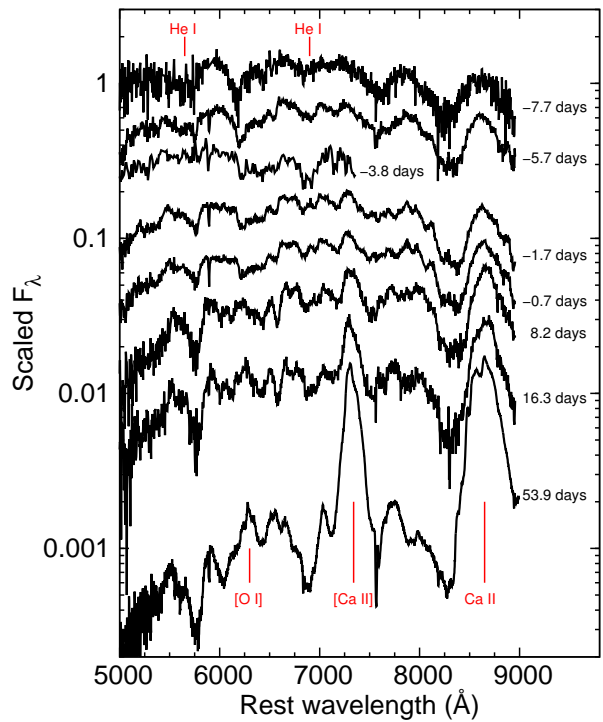}
\caption{The spectral evolution of SN 2019ehk.
The epoch of each spectrum is given with respect to the $R$-band maximum (see Section~\ref{sec:intro}). Only the Galactic extinction has been corrected for.
}
\label{fig:spec}
\end{figure}

The spectra are shown in Figure \ref{fig:spec}, and the log of our spectroscopic observations is shown in Table \ref{table:spec}. We performed optical spectroscopic observations using HOWPol in 6 nights from 5.5 May 2019 ($t=-4.7$~days) to 29.5 May 2019 ($t=30.5$~days).
We used a grism with a spectral resolution of $R\sim 400$ and a spectral coverage of 4,500--9,000~\AA. We observed spectroscopic standard stars in the same nights for the flux calibration.
For the wavelength calibration, we used sky emission lines taken in the object frames. The strong atmospheric absorption bands around 6900 \AA\ and 7600 \AA\ have been removed using the spectra of hot standard stars. 


We also obtained an optical spectrum using the Kyoto Okayama Optical Low-dispersion Spectrograph with an integral field unit \citep[KOOLS-IFU][]{matsubayashi2019} attached to the 3.8-m Seimei telescope of Kyoto University on 9 May 2019 ($t=-3.8$~days).
We used the VPH-blue grism with a wavelength resolution of $R\sim 500$ and a wavelength coverage of 4,000--8,900~\AA. To remove cosmic ray events, we used the L. A. Cosmic pipeline \citep{vandokkum2001}. The data reduction was performed using Hydra package in IRAF \citep{barden1994,barden1995} and a reduction software specifically developed for KOOLS-IFU data. The flux was calibrated using the data of a spectroscopic standard star taken on the same night. For the wavelength calibration, we used arc lamp (Hg and Ne) data.

Another optical spectrum was obtained using the Gemini Multi-Object Spectrograph \citep[GMOS;][]{allington2002,hook2004} attached to the Gemini telescope on 6.3 Jul 2019 ($t \sim 54$~days). We used the $R400$ grism with a slit width of 1~arcsec. The flux calibration was performed using a spectrum of a spectrophotometric standard star taken in the same night. Data reduction was carried out using Gemini IRAF\footnote{https://www.gemini.edu/node/11823}.

\begin{deluxetable*}{lllllll}
\tablewidth{0pt}
\tablecaption{Log of the optical photometry of SN~2019ehk. The Galactic extinction has been corrected for. 
}
\tablehead{
  MJD &  Epoch     & $B$ & $V$ & $R$ & $I$ & Instruments\\
  &  (day)  & (mag) & (mag) & (mag) & (mag) & 
}
\startdata
58604.7 & -11.6 & 16.91(0.11) & 16.05(0.05) & 15.67(0.04) & 15.20(0.03) & HOWPol\\
58605.5 & -10.8 & 16.92(0.04) & 16.05(0.03) & 15.51(0.04) & 15.04(0.03) & HOWPol\\
58606.6 & -9.7 & 17.46(0.03) & 16.43(0.02) & 15.75(0.02) & 15.26(0.02) & HOWPol\\
58607.6 & -8.7 & 18.06(0.18) & 16.69(0.12) & 16.02(0.03) & 15.54(0.04) & HOWPol\\
58608.6 & -7.7 & 18.18(0.03) & 16.86(0.02) & 16.09(0.02) & 15.58(0.02) & HOWPol\\
58609.6 & -6.7 & 18.30(0.11) & 16.77(0.04) & 16.00(0.02) & 15.46(0.02) & HOWPol\\
58610.6 & -5.7 & 18.11(0.05) & 16.64(0.03) & 15.90(0.02) & 15.35(0.02) & HOWPol\\
58612.6 & -3.7 & 18.01(0.04) & 16.51(0.02) & 15.76(0.02) & 15.13(0.02) & HOWPol\\
58614.6 & -1.7 & 18.09(0.07) & 16.46(0.03) & 15.63(0.02) & 14.99(0.02) & HOWPol\\
58615.7 & -0.6 & 18.11(0.05) & 16.45(0.02) & 15.61(0.02) & 14.93(0.02) & HOWPol\\
58619.5 & 3.2 & --- & 16.79(0.05) & 15.74(0.03) & 14.99(0.03) & HOWPol\\
58624.5 & 8.2 & --- & 17.27(0.03) & 16.12(0.03) & 15.28(0.02) & HOWPol\\
58626.6 & 10.3 & --- & 17.50(0.03) & 16.31(0.02) & 15.40(0.02) & HOWPol\\
58628.5 & 12.2 & --- & 17.65(0.03) & 16.48(0.02) & 15.49(0.02) & HOWPol\\
58629.6 & 13.3 & --- & 17.83(0.19) & 16.64(0.14) & 15.52(0.05) & HOWPol\\
58632.5 & 16.2 & --- & 17.85(0.04) & 16.69(0.03) & 15.64(0.02) & HOWPol\\
58634.5 & 18.2 & --- & 17.97(0.06) & 16.83(0.03) & 15.68(0.02) & HOWPol\\
58637.6 & 21.3 & --- & 18.15(0.15) & 16.96(0.04) & 15.68(0.03) & HOWPol\\
58639.5 & 23.2 & --- & 18.21(0.05) & 17.01(0.04) & 15.76(0.03) & HOWPol\\
58646.5 & 30.2 & --- & 18.71(0.1) & 17.30(0.04) & 15.84(0.03) & HOWPol
\enddata
\label{table:opt}
\end{deluxetable*}

\begin{deluxetable}{llllll}
\tablewidth{0pt}
\tablecaption{Log of the NIR photometry of SN~2019ehk.}
\tablehead{
  MJD &  Epoch     & $J$ & $H$ & $K_{\rm s}$ & Instruments\\
  &  (day)     & (mag) & (mag) & (mag)
}
\startdata
58606.6 & -9.7 & 14.80(0.02) & 14.59(0.02) & 14.38(0.03) & HONIR\\
58610.6 & -5.7 & 14.89(0.02) & 14.67(0.02) & 14.42(0.03) & HONIR\\
58612.6 & -3.7 & 14.71(0.03) & 14.41(0.02) & 14.23(0.05) & HONIR\\
58614.6 & -1.7 & 14.53(0.02) & 14.26(0.02) & 14.14(0.03) & HONIR\\
58615.7 & -0.6 & 14.46(0.02) & 14.2(0.02) & 14.08(0.03) & HONIR\\
58617.6 & 1.3 & 14.41(0.02) & 14.11(0.02) & 14.03(0.02) & HONIR\\
58618.5 & 2.2 & 14.40(0.02) & 14.11(0.02) & --- & HONIR\\
58624.5 & 8.2 & 14.76(0.02) & 14.34(0.02) & 14.21(0.02) & HONIR\\
58627.5 & 11.2 & 15.08(0.02) & 14.62(0.02) & 14.46(0.03) & HONIR
\enddata
\label{table:nir}
\end{deluxetable}

\begin{deluxetable*}{lllllll} 
\tablewidth{0pt}
\tablecaption{Log of the spectroscopic observations of SN~2019ehk.}
\tablehead{
 MJD  &  Epoch  & Exposure  &  Coverage  &  Resolution  & Instruments \\
& (day) & (sec)
  }
\startdata
 58608.6 & -7.7 & 3,600 & 4,500--9,000~\AA  & 400 & HOWPol \\
 58610.6 & -5.7 & 2,700 & 4,500--9,000~\AA  & 400 & HOWPol \\
 58612.5 & -3.8 & 3,600 & 4,000--8,900~\AA  & 500 & KOOLS-IFU \\
 58614.6 & -1.7 & 2,700 & 4,500--9,000~\AA  & 400 & HOWPol \\
 58615.6 & -0.7 & 2,700 & 4,500--9,000~\AA  & 400 & HOWPol \\
 58624.5 & 8.2 & 3,600 & 4,500--9,000~\AA  & 400 & HOWPol \\
 58632.6 & 16.3 & 3,600 & 4,500--9,000~\AA  & 400 & HOWPol \\
 58670.3 & 53.9 &1,800 & 5,200--9,800~\AA & 400 & GMOS
\enddata
\label{table:spec}
\end{deluxetable*}

\section{Results}
\label{sec:result}
\subsection{Spectra}
\label{sec:spec}

Figure \ref{fig:spec} shows the spectral evolution of SN~2019ehk from the end of the first peak ($t=-7.7$~days) to the tail phase ($t=53.9$~days). 
Several absorption lines are identified in the spectra before the second maximum ($t<0$~days); He~\si $\lambda$5876, He~\si $\lambda$ 6678, Si~\sii $\lambda$ 6355, O~\si $\lambda$ 7774, and Ca~\sii NIR triplet. 
Overall features indicate that SN 2019ehk is a member of SNe Ib according to the standard criteria. The strong Na~\si~D narrow absorption line indicates that the absorption within the host galaxy is substantial (Section~\ref{sec:extinction} for further details). After the second peak ($t>0$~days), spectral features quickly evolve. Prominent Ca~\sii and \caii emission lines are developed already at $t=16.3$ days, showing a rapid transition to the nebular phase. These features, a rapid evolution to the nebular phase and the prominent Ca emissions, are characteristics of a Ca-rich transient.

\begin{figure}[h]
\centering
\includegraphics[width=8cm,clip]{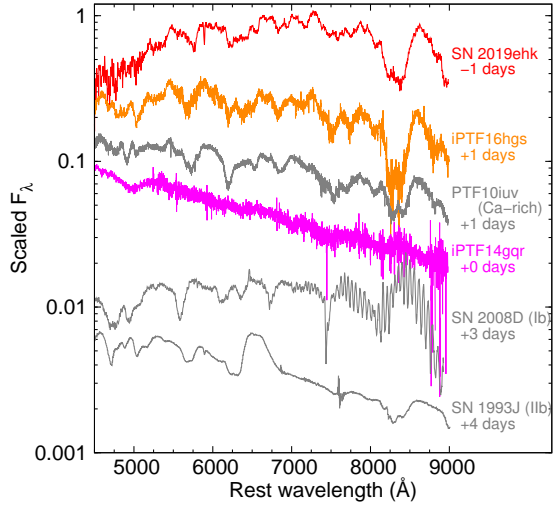}
\caption{The Spectrum of SN 2019ehk around the second peak ($t \sim 0$~days) compared  
with the following objects; SN Ib 2008D \citep{modjaz2009}, SN IIb 1993J 
\citep{barbon1995}, iPTF16hgs \citep{de2018}, iPTF14gqr \citep{de2018_14ft}, and (canonical) Ca-rich transient PTF10iuv \citep{kasliwal2012}, at similar epochs.
Only the Galactic extinction has been corrected for in SN~2019ehk.
}
\label{fig:com_sp}
\end{figure}

In Figure \ref{fig:com_sp}, we compare SN~2019ehk with SN Ib 2008D \citep{modjaz2009}, SN IIb 1993J \citep{barbon1995}, iPTF14gqr \citep{de2018_14ft}, iPTF16hgs \citep{de2018}, and Ca-rich transient PTF10iuv \citep{kasliwal2012} at $t \sim 0$ days (i.e., around the maximum brightness). 
Overall, the spectrum of SN~2019ehk is similar to those of SN Ib 2008D and the Ca-rich transients (iPTF16hgs and PTF10iuv). Indeed, there are no significant differences between the Ca-rich transients 
and (some of) SNe Ib at this stage. 
At a closer look, the line profiles of He~\sib, O~\sib, and Fe~\sii seen in 
SN 2019ehk show a better match to those of the Ca-rich transients than SN 2008D. 
The spectrum of iPTF14gqr in this early phase is distinct from the Ca-rich transients, confirming the argument by \citet{de2018_14ft} that it is not classified as a `canonical' Ca-rich transient. 

\begin{figure}[h]
\centering
\includegraphics[width=8cm,clip]{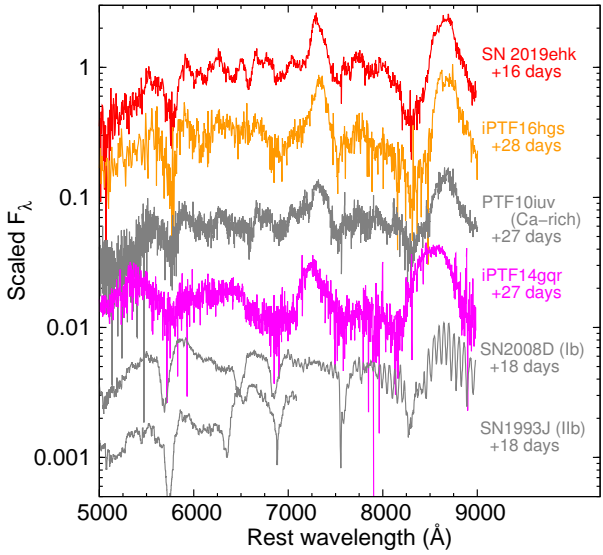}
\caption{Same as Figure~\ref{fig:com_sp}, but at $t \sim 20$~days. 
}
\label{fig:com_ca20}
\end{figure}

Emission lines of Ca~\sii and \caii become prominent for iPTF16hgs and PTF10iuv at $t \sim 20$ days (Figure~\ref{fig:com_ca20}), which is the definition of the `Ca-rich' transient class \citep{kawabata2010,perets2010}.
SN 2019ehk shows a  strikingly similar spectrum. 
While the absorption lines of He~\si are still visible, the spectra of the Ca-rich transients are distinct from SNe Ib at this phase, and thus SN 2019ehk should be definitely classified as a Ca-rich transient, not as a canonical SN Ib. 
Interestingly, iPTF14gqr becomes to resemble the other Ca-rich transients at this phase, showing strong \caii and Ca~\sii NIR triplet. Therefore, iPTF14gqr can also be classified as a Ca-rich transient according to the standard criterion. We however note that iPTF14gqr shows broader Ca emissions than the others, and some lines identified in the other Ca-rich transients are not visible in its spectrum. As such, iPTF14gqr could be classified as a peculiar Ca-rich transient, which might indicate that its origin is different from the canonical Ca-rich transients. 

\begin{figure}[h]
\centering
\includegraphics[width=8cm,clip]{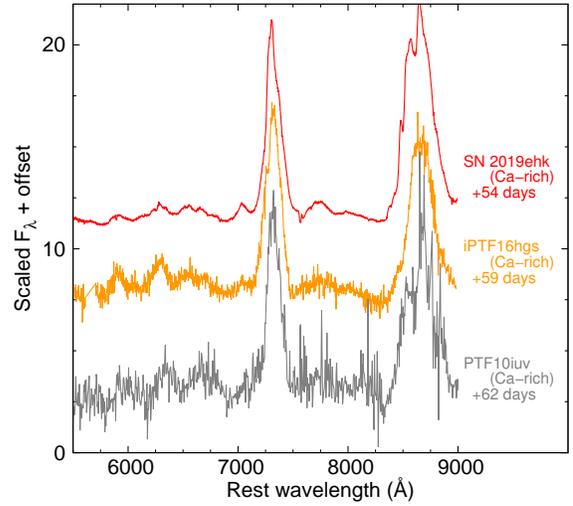}
\caption{
The spectrum of SN 2019ehk at a late phase ($t = 54$~days) as compared with those of Ca-rich transients 
iPTF16hgs \citep{de2018} and PTF10iuv \citep{kasliwal2012} at similar epochs.
Only the Galactic extinction has been corrected for in SN~2019ehk.
}
\label{fig:com_ca55}
\end{figure}

The difference between (normal) SNe Ib and Ca-rich transients becomes more obvious in the later phases. All the Ca-rich transients show strong \caii and Ca~\sii emission lines, and weak or no \oi at $t\sim 55$~days (Figure~\ref{fig:com_ca55}). 
The velocity seen in SN~2019ehk, as measured from the full-width-half-maximum (FWHM) of a Gaussian function fitted to the \caii profile, is $\sim$5300~\kms. This is again similar to other (canonical) Ca-rich transients \citep{kasliwal2012} (e.g., as compared to PTF10iuv).
The \oi seen in SN~2019ehk is very weak even among Ca-rich transients.

\begin{figure}[h]
\centering
\includegraphics[width=8cm,clip]{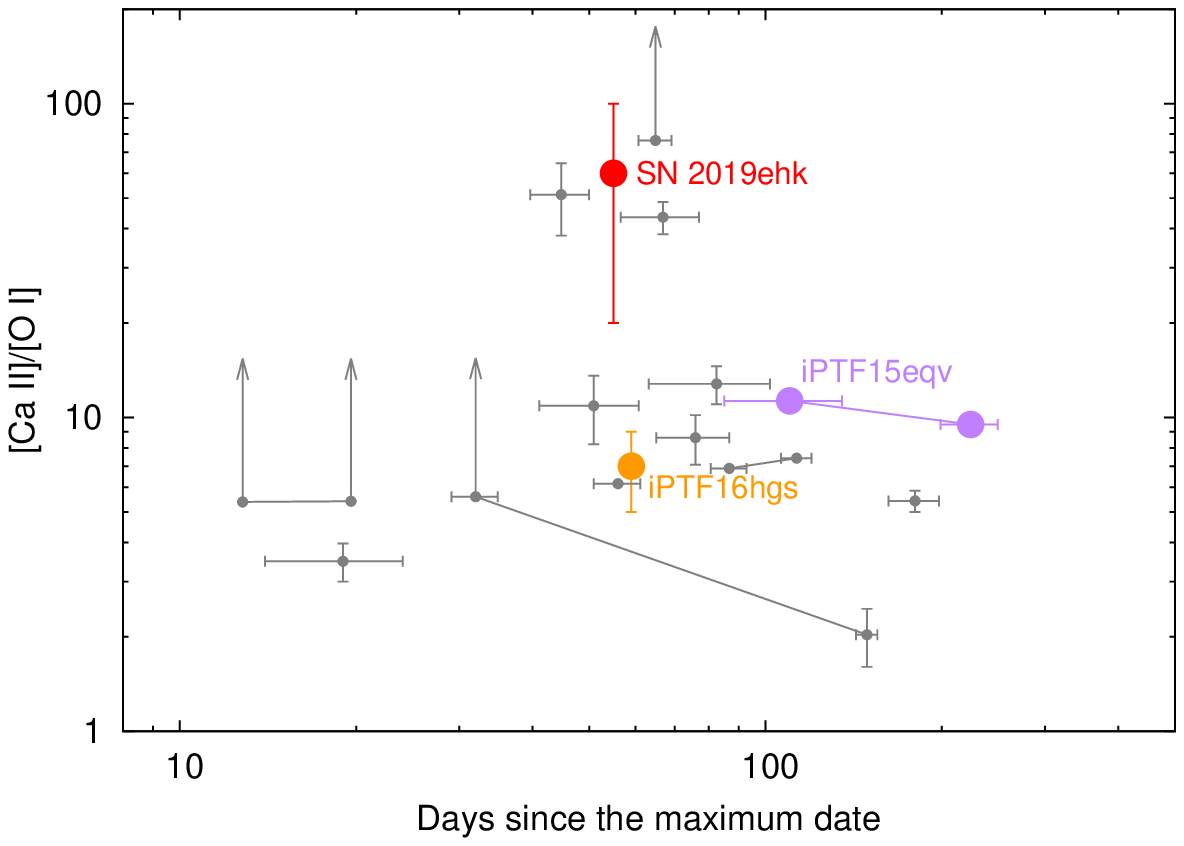}
\caption{The ratio of \cao compared with those for other Ca-rich transients, including iPTF16hgs \citep{prentice2019,milisavljevic2017,de2018}.
}
\label{fig:OCa}
\end{figure}

Figure~\ref{fig:OCa} shows the evolution of the \cao ratio as compared with those measured from spectra of a sample of Ca-rich transients presented by \citet{valenti2014}. The \cao ratio of SN~2019ehk at $t \sim 54$~days is measured to be $60 \pm 40$, which is within the range generally seen in the Ca-rich transients, and is among the largest. 

\begin{figure}[h]
\centering
\includegraphics[width=8cm,clip]{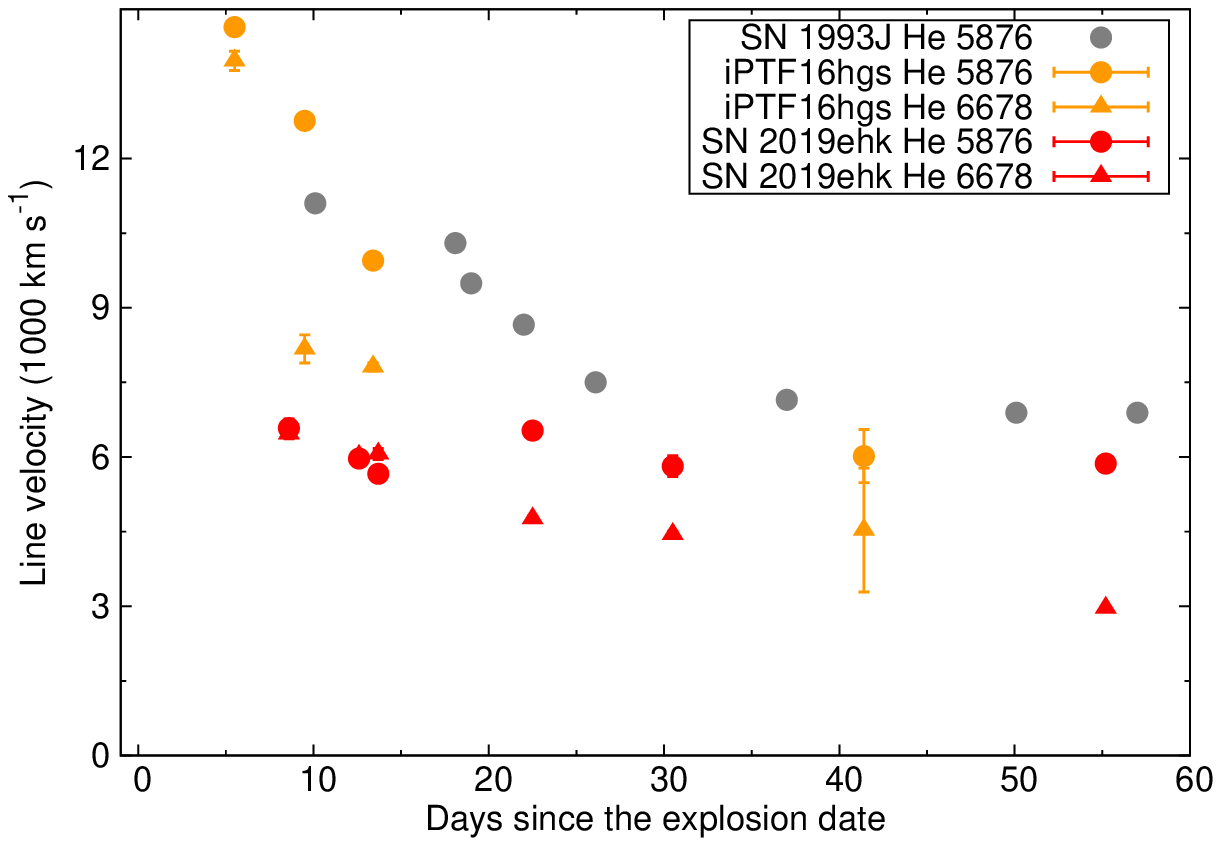}
\caption{The He~\si $\lambda$5876 and $\lambda$6678 velocity evolution of SN~2019ehk as compared with those of SN IIb 1993J \citep{barbon1995} and iPTF16hgs \citep{de2018}. 
The explosion date is estimated as $t=-14.4~days$.
}
\label{fig:he}
\end{figure}

Figure~\ref{fig:he} shows the evolution of the He~\si line velocities of SN~2019ehk. The velocities are derived from a Gaussian fit to the He~\si $\lambda$5876 and He~\si $\lambda$6678 absorption line profiles. We do not see clear evolution in the velocity of He~\si $\lambda$ 5876, but it is likely contaminated by Na~\si~D. The velocity measured for He~\si $\lambda$ 6678 is around 6500~\kms at $t \sim -5$~days,
and declines to $\sim$3,000~\kms at $t \sim 54$~days. The velocity and its evolution here roughly overlap with those of iPTF16hgs. 

The spectral features and their evolution show that SN 2019ehk clearly belongs to the Ca-rich transient class.
It shows very similar spectral properties with those of iPTF16hgs, and these two objects also share striking similarities in the light curves (see Section \ref{sec:lc}). iPTF14gqr is distinct from SN 2019ehk, iPTF16hgs,
and other Ca-rich transients in the spectral evolution, while its similarities in the late phase spectrum and in the light curve properties (Section~\ref{sec:lc}) to SN 2019ehk and iPTF16hgs suggest a link among these three objects.

\subsection{Extinction correction}
\label{sec:extinction}
 
The Galactic extinction is negligibly small for SN 2019ehk \citep[\ebv$=0.026$ mag;][]{schlafly2011}. However, the extinction within the host galaxy along the line-of-sight toward SN 2019ehk is highly uncertain. Our spectra show a very deep Na~\sib~D absorption line, indicating substantial extinction within the host galaxy (see Figure~\ref{fig:spec}). We measure the equivalent width (EW) as $3.2 \pm 0.3$ \AA; 
this is beyond the applicability of the extinction-EW relation suggested by \citet{poznanski2010}. 

\begin{figure}[h]
\centering
\includegraphics[width=8cm,clip]{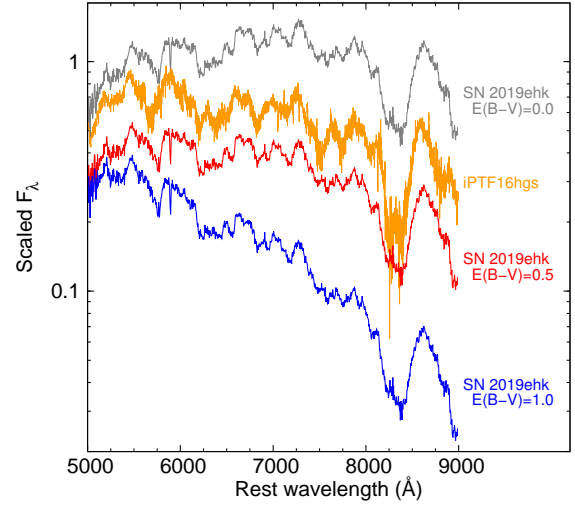}
\caption{The extinction-corrected spectra of SN~2019ehk as compared with iPTF16hgs, with different values assumed for the host extinction: \ebv$=0.5$ (red) and $1.0$ (blue). The gray line shows the original spectrum of SN~2019ehk at $t=-1$~days, and the green line shows the spectrum of iPTF16hgs at $t=1$~days.
}
\label{fig:19ehk_16hgs}
\end{figure}

Given the similarities between SN 2019ehk and iPTF16hgs, we could match the spectra of these two objects to constrain the extinction. Figure~\ref{fig:19ehk_16hgs} shows the spectral comparison at $t \sim 0$~days,
with different values of \ebv applied to SN 2019ehk (\ebv$=0.5$ and $1.0$ mag). With \ebv$=0.5$ mag, the two spectra match quite well. The black body temperature of SN~2019ehk is $\sim$5,400~K and $\sim$8,100~K assuming the extinction of \ebv$=0.5$ and $1.0$ mag, respectively. Since the spectral features of SN~2019ehk are similar to those of iPTF16hgs, the temperature should not be very different between the two objects; if the temperature would be different by $\gsim$50\%, difference in the spectral features would likely become significant \citep{Nugent1995}. This places a rough upper limit of \ebv$\sim1.0$ mag for SN~2019ehk. 

\begin{figure}[t]
\centering
\includegraphics[width=8cm,clip]{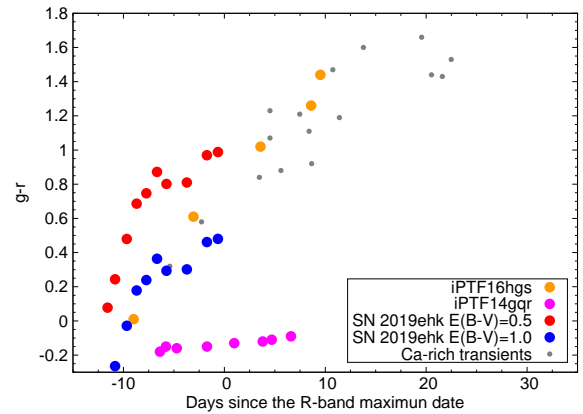}
\caption{The $g-r$ color evolution of SN~2019ehk as compared with iPTF14gqr, iPTF16hgs, and other Ca-rich transients (PTF10iuv, PTF11kmb, PTF12bho). The red and blue filled circles denote the color of SN 2019ehk corrected for the host extinction, assuming \ebv$=0.5$ and \ebv$=1.0$ mag, respectively.
The SDSS magnitudes are given in the AB system.
}
\label{fig:color}
\end{figure}

To compare the color evolution of SN~2019ehk and other Ca-rich transients,
we convert the magnitudes of SN~2019ehk to the SDSS system using the relations given by \citet{jordi2006}.
In the comparison using the SDSS system, we apply the AB magnitude.
We also confirm that these SDSS (AB) magnitudes are consistent with those estimated from the spectra of SN~2019ehk.
Figure \ref{fig:color} shows the $g-r$ color evolution of SN~2019ehk as compared to iPTF16hgs, iPTF14gqr, and other Ca-rich transients.
The color evolution of SN~2019ehk roughly matches to those of iPTF16hgs and other Ca-rich transients, for \ebv between 0.5 and 1.0 mag.
This range is consistent with that from the spectral matching to iPTF16hgs.
Therefore, we estimate the host extinction to be between \ebv$=0.5$ and $1.0$ mag.

An issue to obtain the absolute magnitudes is a value of $R_v$.
Usually the Galactic value ($R_v=3.1$) is adopted for CCSNe,
while it is suggested to be lower ($R_v \sim 2$) for SNe Ia \citep{wang2008}.
The host extinction is usually negligible for Ca-rich transients,
thus the typical value for Ca-rich transients is uncertain.
In this paper, we adopt the Galactic value, \ie $R_v=3.1$,
which is more likely the case than the smaller value given our interpretation of SN 2019ehk,
independent from $R_v$, as a member of CCSNe (see Section \ref{sec:discussion}).
While adopting $R_v \sim 2$ will reduce the estimated absolute magnitude by $ \sim 0.5-1.0$ mag,
this would not affect our main conclusions since we already consider a range of \ebv$= 0.5-1.0$ mag.

\subsection{Light curves}
\label{sec:lc}

Figure~\ref{fig:lc} shows our multi-band light curves. Note that these magnitudes are corrected only for the Galactic extinction (see Section~\ref{sec:extinction}). The light curve shows a clear rise to the first peak in the optical bands after its discovery. The similar initial emission was detected also for iPTF14gqr and iPTF16hgs,
while the rising part toward the first peak was missed for both of them. This early emission has not been seen in the other Ca-rich transients. The first peak is reached at $t=-10.8$~days for SN~2019ehk.

The decline rate after the second peak (between $t=-0.6$ and $t=13.3$~days) is 0.09, 0.07, and 0.05~\magdays in the $V$, $R$, and $I$ bands, respectively. After $t \sim 13.3$~days, the decline becomes slower. The decline rate between $t=26$ and $38$ days is 0.05, 0.05, and 0.03 \magdays in the $V$, $R$, and $I$ bands, respectively. A bluer band shows a steeper decline.



\begin{figure}[t]
\includegraphics[width=8cm,clip]{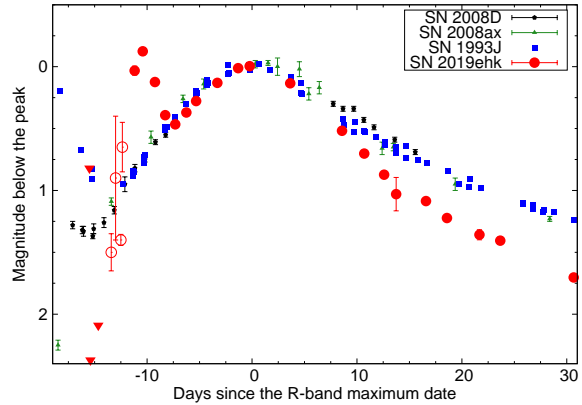}
\caption{The $R$-band light curve of SN 2019ehk compared with those of well-observed SNe: SN Ib 2008D \citep{modjaz2009}, SN IIb 2008ax \citep{pastorello2008}, SN IIb 1993J \citep{richmond1996_93J}. All the magnitudes are relative to the peak.
}
\label{fig:abs_normal}
\end{figure}

\begin{figure}[t]
\includegraphics[width=8cm,clip]{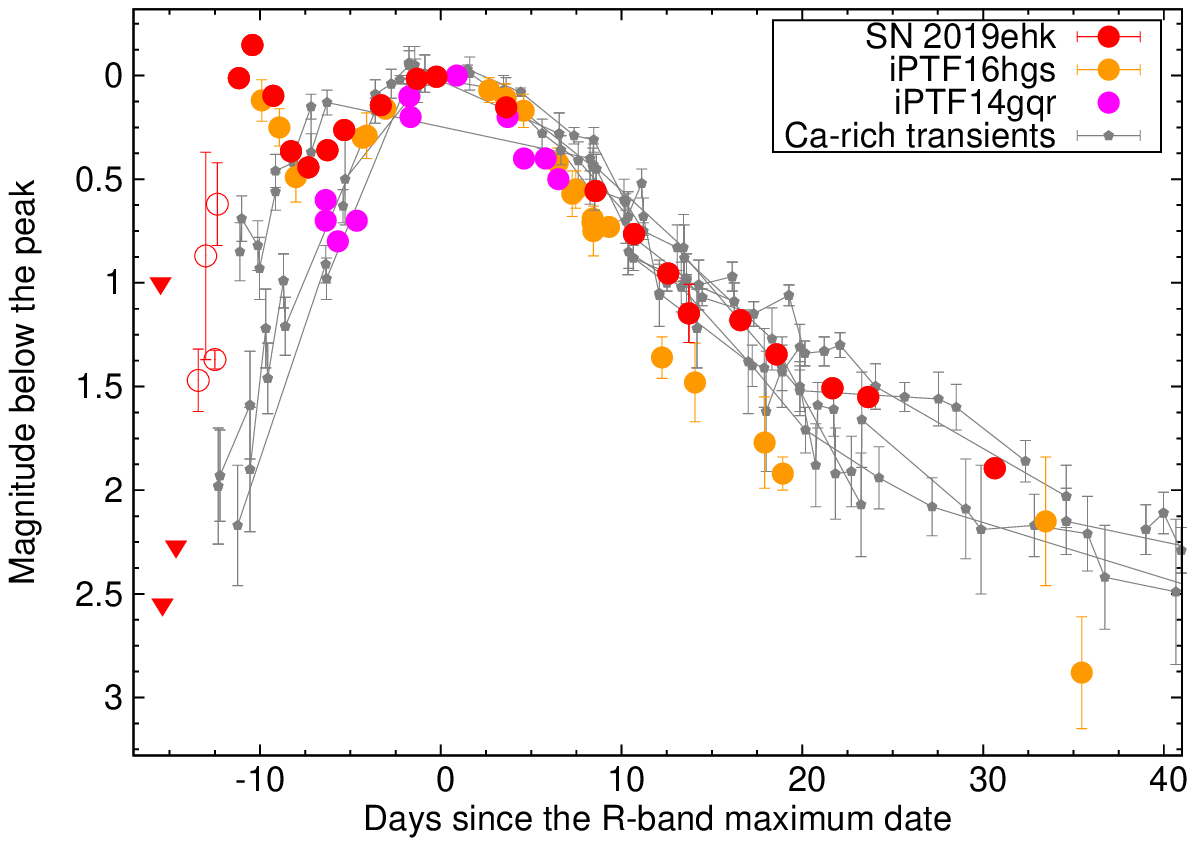}
\includegraphics[width=8cm,clip]{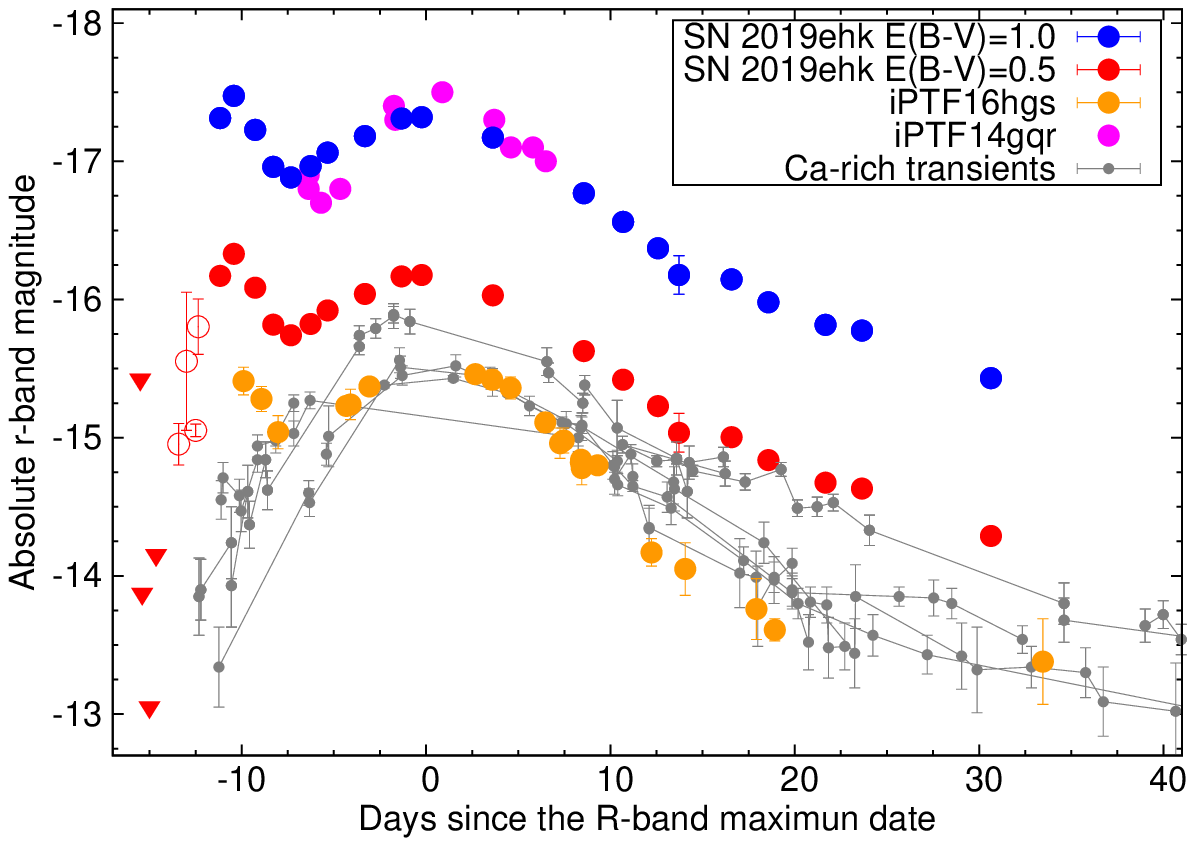}
\includegraphics[width=8cm,clip]{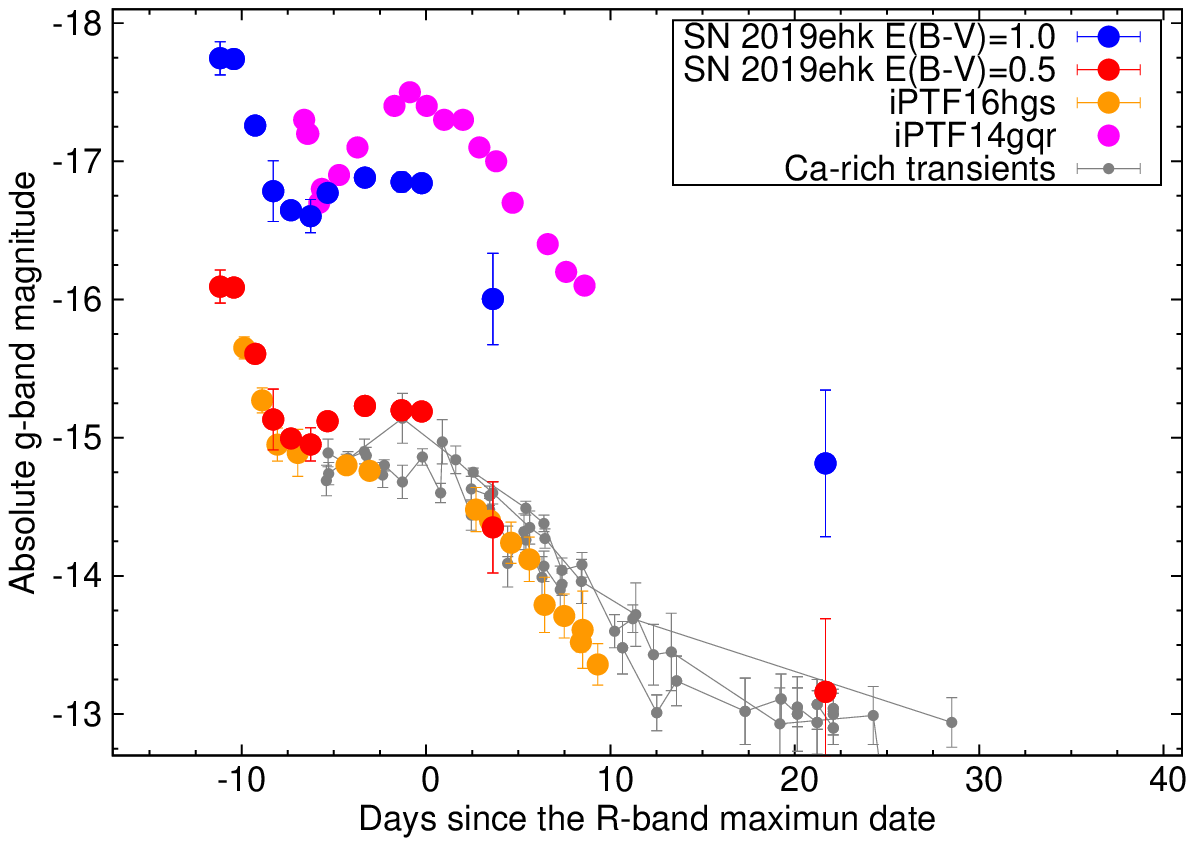}
\caption{(Top) The $r$-band light curve of SN 2019ehk compared with iPTF14gqr \citep{de2018_14ft}, iPTF16hgs \citep{de2018}, and other Ca-rich transients PTF10iuv, PTF11kmb, PTF12bho \citep{kasliwal2012,lunnan2017}.
(Middle and Bottom) Same as the top panel but the magnitudes are shown in the absolute magnitude scale, in the $r$-band (middle) and in the $g$-band (bottom), given in the AB system. 
The light curve of SN~2019ehk is shown for two different values of the host extinction; \ebv$=0.5$ mag (red filled circle) and  \ebv$=1.0$ mag (blue one).
The magnitudes of SN~2019ehk are given in the AB magnitude.
In these figures, $R_v=3.1$ is assumed.
}
\label{fig:abs_ca}
\end{figure}

We compare the $R$-band light curve of SN 2019ehk with those of canonical SNe IIb and Ib in Figure~\ref{fig:abs_normal}.
The same comparison, but for the $r$ band (see Section~\ref{sec:extinction}), is shown with iPTF14gqr, iPTF16hgs, and other Ca-rich transients in the top panel of Figure~\ref{fig:abs_ca}.
In these figures, we also plot the magnitudes of SN~2019ehk reported by other sources (see Section~\ref{sec:obs}).
These early points are shown here only for demonstration purpose, given the differences in the band passes even though the central wavelengths are close to the $R$ or $r$ band.
For the comparison, we shift the light curves vertically to match to the peak magnitude of SN 2019ehk, and horizontally to match to the peak date. 


The middle and bottom panels of Figure \ref{fig:abs_ca} show the absolute magnitude light curves of SN~2019ehk in the $r$ and $g$ bands, respectively, as compared with those of iPTF14gqr, iPTF16hgs, and other (canonical) Ca-rich transients (see Section~\ref{sec:extinction}).
To show the uncertainty associated with the extinction, the absolute magnitudes of SN~2019ehk are corrected for the host excitation with \ebv$=0.5$ mag (red filled circles) or \ebv$=1.0$ mag (blue). 

Figure~\ref{fig:abs_normal} shows that the light curve evolution is much faster for SN 2019ehk than the well-studied SNe IIb and Ib.
It is especially clear in terms of the rising time to the second peak since the (estimated) explosion date; 22.7, 20.0, 23.6, and 12.7~days for SNe 1993J, 2008D, 2008ax, and 2019ehk, respectively.
The same tendency is also seen in the decline rate after the second maximum; 0.045, 0.05, 0.055, and 0.07~\magdays in SNe 1993J, 2008D, 2008ax, and 2019ehk, respectively. 

SN~2019ehk, iPTF16hgs, and other Ca-rich transients show similar rising time as seen in Figure \ref{fig:abs_ca}.
For example, the rising time and the decline rate for iPTF16hgs in the $r$ band are 12.7 days and 0.10~\magdays, respectively, which are both similar to those of SN 2019ehk. iPTF14gqr shows faster rise and fall than these objects, suggesting a smaller amount of ejecta.
In any case, the light curve evolution of iPTF14gqr is qualitatively similar to SN 2019ehk and the Ca-rich transients, being much faster than canonical SNe Ib.  

The decline rate of SN~2019ehk after $t\sim10$~days is within a range observed for the Ca-rich transients, although it is on the slowest side.
On the other hand, iPTF16hgs shows the fastest decline among the Ca-rich transients.
If SN~2019ehk and iPTF16hgs belong to the same subpopulation within the Ca-rich transient class (as argued in this paper, along with iPTF14gqr), it might suggest that this subpopulation has much more diverse properties than the other (main) population within the Ca-rich transient class. 



The $r$-band peak magnitude of SN~2019ehk at $t\sim0$~days is by $\sim$1~mag brighter than those of PTF10iuv and iPTF16hgs in the case of \ebv$=0.5$ mag
\footnote{Only if we would adopt a combination of \ebv $\sim 0.5$ mag and $R_v \sim 2$ mag for the host extinction 
(\ie the assumptions leading to the minimal amount of the extinction),
the absolute magnitude of SN 2019ehk would be similar to iPTF16hgs.}.
Even taking the uncertainty in the host extinction into account, SN~2019ehk is much more luminous in its (second) peak than a bulk of the Ca-rich transients.
It can be comparable to the peak luminosity of iPTF14gqr, which also shows a distinctly brighter peak than the other comparison objects. 

The first peak is reminiscent of early emission from some CCSNe, exemplified by SN IIb 1993J.
The magnitude of this emission relative to the second-peak magnitude, as well as the declining rate (and the duration from the estimated explosion date), are indeed similar between SN 1993J and SN 2019ehk.
The emission is well understood as the `cooling envelope emission' for the case of SN 1993J, with its extended H-rich progenitor \citep{richmond1996_93J}.
The similar early emission is also seen for iPTF14gqr and iPTF16hgs, while such a feature has not been detected for the other (canonical) Ca-rich transients.
The existence of the first peak would suggest that SN~2019ehk, iPTF14gqr, and iPTF16hgs may form a subpopulation within the Ca-rich transient class (or, a population of explosions which resemble the canonical Ca-rich transients in the observational features).
The similarity in the properties of the first peak may further suggest that they are related to the death of massive stars (Section~\ref{sec:bump} for further details). 



\section{Discussion}
\label{sec:discussion}




Based on the observational data presented in Section~\ref{sec:result}, we discuss properties and a possible origin of SN 2019ehk. The discussion is also given for a possible relation of SN 2019ehk to iPTF14gqr, iPTF16hgs, and the Ca-rich transients in general. In Section~\ref{sec:peak}, we estimate the  properties of the ejecta from the  observational properties around the second peak. The origin of the first peak is discussed in Section~\ref{sec:bump}. In Section~\ref{sec:progenitor}, we argue that SN 2019ehk is an explosion of a massive star, likely an USSN. In Section~\ref{sec:pop}, we discuss a subpopulation within the Ca-rich transient class including SN~2019ehk.


\subsection{Properties of the ejecta and implications for the progenitor}
\label{sec:peak}
First, we construct a quasi-bolometric light curve of SN~2019ehk by integrating the spectral energy distribution (SED) in the $B$, $V$, $R$ and $I$ bands. The integration is performed by interpolating the SED by trapezium functions. The conversion from the observed magnitude into the flux is carried out using the filter function \citep{bessell1990} and the zero-point flux in each band.

\begin{figure}[h]
\centering
\includegraphics[width=8cm,clip]{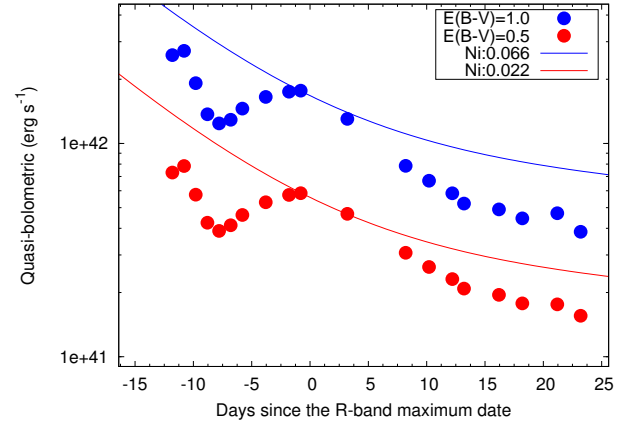}
\caption{The bolometric light curves(s) of SN~2019ehk as compared to the ${}^{56}$Ni/Co/Fe decay full-trapping model light curve(s) \citep{nadyozhin1994} with different assumptions on the host extinction; \ebv$=0.5$ (red circles and line) and \ebv$=1.0$ mag (blue), and $R_v = 3.1$.
The explosion date is estimated as $t=-14.4~days$.
}

\label{fig:bol}
\end{figure}

We show the bolometric light curve in Figure~\ref{fig:bol}.
The bolometric light curve is constructed assuming two different values of the host extinction,
\ebv$=0.5$ and $1.0$ mag (see Section~\ref{sec:extinction}).
The ratio of the optical flux (in the $B$, $V$, $R$, and $I$ bands) to the total flux is assumed to be 60\% \citep{drout2014}.
The first peak of SN~2019ehk is also confirmed in the quasi-bolometric light curve. The second-peak luminosity, $L_{{\rm peak}}$, is $\sim 6 \times 10^{41}$ \ergs in the case of the low extinction (\ebv$=0.5$ mag), which already exceeds a typical range found for the canonical Ca-rich transients \citep{lunnan2017,kasliwal2012}. It would even be as bright as $L_{\rm peak} \sim 2 \times 10^{42}$ \ergs if \ebv$=1.0$ mag, being comparable to the peak luminosity of iPTF14gqr. The rise time to the second peak is estimated to be 14.4~days, which is similar to iPTF16hgs \citep[12.6~days;][]{de2018} and other Ca-rich transients.

We calculate the \Nifs mass, \MNifs, as $0.022-0.066$~\Msun from the second-peak luminosity  \citep{arnett1982}, where the range reflects the extinction uncertainties.
Even with \ebv$=0.5$ mag, this is significantly larger than the values found for most of Ca-rich transients \citep{lunnan2017,kasliwal2012} including iPTF16hgs \citep{de2018}. 
The estimated \Nifs mass is between iPTF16hgs and iPTF14gqr, or similar to iPTF14gqr, depending on the treatment of the host extinction toward SN~2019ehk
\footnote{\MNifs of SN~2019ehk can be similar to that of iPTF16hgs only for the (unlikely) combination of \ebv$\sim 0.5$ mag and $R_v \sim 2$ for the host extinction.}. 

We estimate the kinetic energy and the ejecta mass for SN 2019ehk from the analysis of the light-curve evolution.
We fit the $r$-band light curve of SN 2019ehk to that of iPTF16hgs by stretching the light curve width around the second maximum. We then obtain the ratio of the timescale of SN 2019ehk to iPTF16hgs as $\sim1.2$.

Figure~\ref{fig:he} suggests that the expansion velocities are similar between SN 2019ehk and iPTF16hgs, with the ratio of $\sim 0.8$ between two objects.We further test this by the overall spectral property. After bending the second-maximum ($t\sim0$~days) spectrum of SN 2019ehk to match to the continuum of iPTF16hgs (Figure~\ref{fig:19ehk_16hgs}),
we introduce an artificial shift in the wavelength to the spectrum of SN 2019ehk, to find the best match in the positions of the absorption minima of different lines.
We thereby find that the ratio of the line velocities of SN 2019ehk to iPTF16hgs is on average $v_{\rm 19ehk}$/$v_{\rm 16hgs}\sim 0.8$, being consistent with the estimate using the He velocity. 


We then convert these ratios in the light curve widths and the expansion velocities to the ejecta mass ($M_\ej$) and the kinetic energy of the ejecta ($\ek$), using the following relations; 
\begin{equation}
\tau \propto M_\ej^{3/4} \cdot \ek^{-1/4},
\end{equation}
\begin{equation}
v_\ej \propto M_\ej^{-1/2} \cdot \ek^{1/2}.
\end{equation}

For the normalization in this scaling method, we use the ejecta properties estimated for iPTF16hgs \citep[$M_\ej= 0.38~\Msun$ and $\ek= 2.3 \times 10^{50}$~erg;][]{de2018}.
We then obtain $M_\ej= 0.43$~\Msun and $\ek= 1.7 \times 10^{50}$~erg for SN~2019ehk.

These properties are similar to those obtained for iPTF16hgs (and the other Ca-rich transients in general), as expected from the similar properties seen in their light curves and spectra. The ejecta mass is significantly larger than that of iPTF14gqr while the kinetic energy is similar \citep[$M_\ej= 0.2~\Msun$ and $\ek= 2 \times 10^{50}$~erg;][]{de2018_14ft}; these properties are consistent with the slower evolution and lower line velocities in SN 2019ehk than in iPTF14gqr. 

\subsection{Analysis of the first peak}
\label{sec:bump}
Here, we investigate an origin of the first peak in the light curves. 
Given the large uncertainty in the host extinction, we consider a range of the possible extinction,
between \ebv$=0.5$ and $1.0$ mag as roughly constrained by the color at the second peak (see Section \ref{sec:extinction}).
%



The first peak observed for SN~2019ehk is reminiscent of the so-called cooling-envelope emission frequently observed for CCSNe with an extended envelope (\eg see SN IIb 1993J in Figure~\ref{fig:abs_normal}),
where the thermal energy deposited by the propagation of the shock wave is diffused out in the cooling time scale. The mechanism is indeed not limited to a `stellar envelope' but is applicable to a `confined' CSM as long as the CSM is already swept up by the ejecta and behaves like a cooling envelope. 

For the analysis of this process, we follow the formalism presented by \citet{piro2015}. In doing this, we use a semi-analytic code developed by \citet{maeda2018} to simulate a similar emission process. We calibrated the code by comparing the output to the results of \citet{piro2015} for the same parameter set. Indeed, \citet{de2018} used the same formalism to analyze the first peak seen in iPTF16hgs.
We have also confirmed that we obtain a consistent result with \citet{de2018} if we use the same input parameters; this guarantees a fair comparison between the natures of SN~2019ehk and iPTF16hgs. 

As the input parameters, we adopt $M_\ej= 0.43$~\Msun for the ejecta mass and $\ek = 1.7 \times 10^{50}$~erg for the kinetic energy (see Section~\ref{sec:peak}). 
The luminosity is sensitive to the radius of the envelope (or CSM) while the diffusion time scale is so to the envelope (CSM) mass, and there is no degeneracy between these two parameters to derive. We first try to obtain a rough match to the $V$-band light curve, and the same model is adopted to generate the $B$ and $R$-band light curves. 

\begin{figure}[t]
\includegraphics[width=8cm,clip]{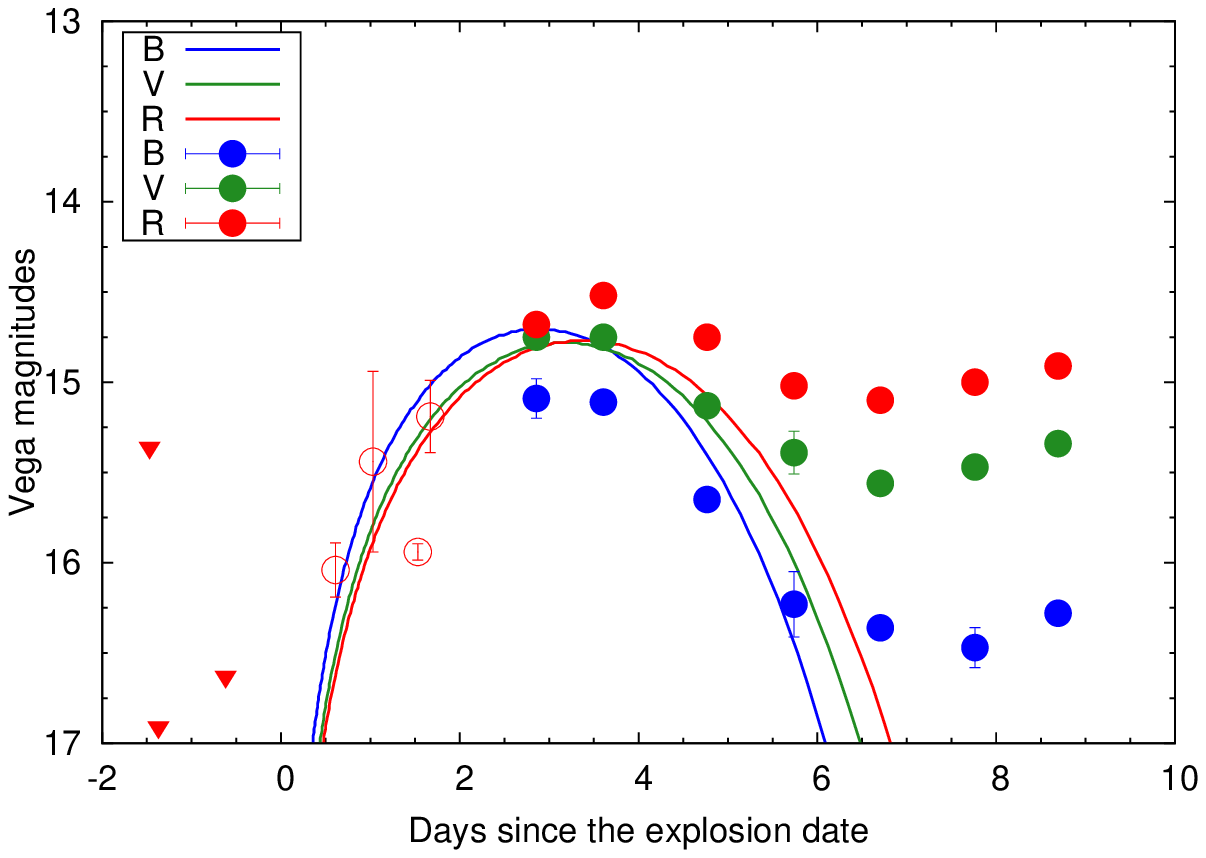}
\includegraphics[width=8cm,clip]{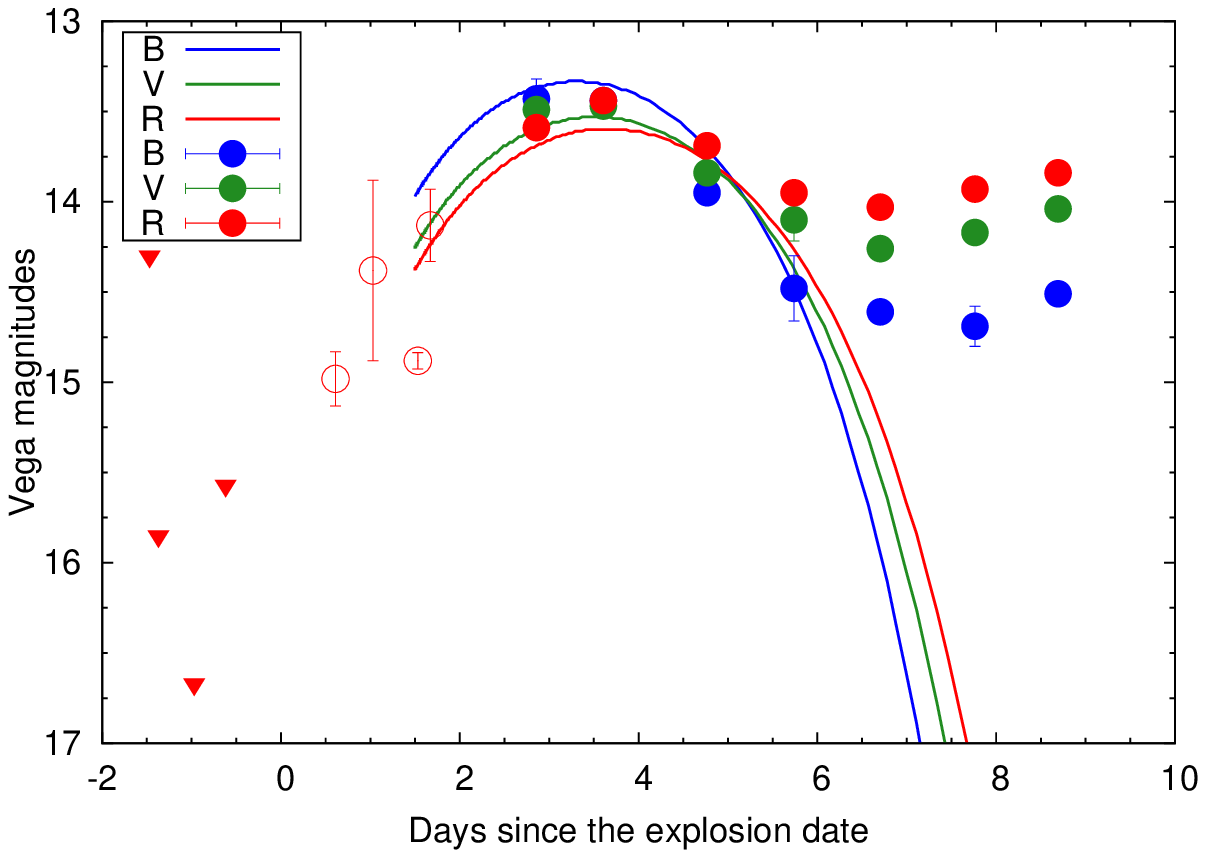}
\caption{The $B$, $V$, and $R$-band light curves of SN~2019ehk and the cooling-envelope/CSM models that fit to the $V$-band light curve. The host extinction is assumed to be \ebv$=0.5$ (top) or \ebv$=1.0$ mag (bottom).
The explosion date is estimated as $t=-14.4~days$.
}
\label{fig:bump}
\end{figure}

As shown in Figure~\ref{fig:bump}, we can obtain a reasonable fit to the early $V$-band light curve. The envelope/CSM mass is $\sim 0.04-0.05$~\Msun irrespective of the extinction assumption. The outer radius of the envelope/CSM is sensitive to the assumed extinction: 130~\Rsun, 300~\Rsun, and 1,500~\Rsun for \ebv$=0.5$, $0.7$, and $1.0$ mag, respectively. While the mass is similar to that derived for iPTF16hgs \citep{de2018}, the radius required for SN~2019ehk is substantially larger. This is a result from the brighter first peak luminosity ($L_{{\rm bump}}$) in SN~2019ehk even for \ebv$=0.5$ mag. 

We note that the expected model color is far too blue to be consistent with the observations of SN~2019ehk, as long as \ebv $< 1.0$ mag. The cooling-envelope emission predicts that the $V$-band peak is realized when the photospheric temperature decreases to $\sim 10,000 - 20,000$ K, as an outcome of the shift of the blackbody peak from the UV to the optical bands. However, the observed color of the first peak is red ($V-R > 0$ mag) if \ebv$<1.0$ mag.
There is then no solution to explain the color by the cooling-envelope emission under the low-extinction assumption (Figure~\ref{fig:bump}). 

The physically reasonable solution with the cooling-envelope emission model can be found only if we assume a high value of the extinction (\ebv$=1.0$ mag; see Figure~\ref{fig:bump}). The derived radius is as large as $\sim 1,500$~\Rsun. This is even comparable to one of the largest SN IIP progenitors discovered through the pre-explosion image analysis \citep{huang2018}. Our spectra do not exhibit any hydrogen features,
and such a large and He-rich `envelope' is physically not realized. On the other hand, the mass and radius here are very similar to those suggested for the `confined' CSM around CCSNe commonly inferred through the flash spectroscopy \citep{galyam2014,yaron2017} or early light curve behaviors \citep{moriya2017,moriya2018,morozova2017,morozova2018,forster2018}.
We suggest that this can be interpreted as the existence of the confined CSM around SN~2019ehk, which might further support its origin as a core collapse of a massive star. The corresponding mass loss rate is $\sim 1.5$~\Msun~\years assuming the mass-loss velocity of $1,000$ km s$^{-1}$.
This is high due to the large wind velocity assumed here, but it involves a large uncertainty even more than an order of magnitude.


As clarified above, another emission mechanism is required if \ebv$< 1.0$ mag. An alternative mechanism is the situation in which the ejecta are still propagating within the optically-thick CSM at the first peak of the light curve. This allows a larger photospheric radius and a lower temperature, and thus a redder color.
If \ebv$=0.5$ mag, the blackbody fit to the SED at the first peak suggests the photospheric temperature of $\sim 7,000$~K. The observed luminosity and temperature for the case of \ebv$=0.5$ mag indicate the black body radius of $\sim 11,000$~\Rsun (i.e., $\sim 8 \times 10^{14}$ cm). Similarly, if we assume \ebv$=0.7$ mag, then the photospheric temperature and the radius are $\sim 8,500$~K and $\sim 9,400$~\Rsun ($\sim 7 \times 10^{14}$ cm), respectively. This is again within a range suggested for the confined CSM. We can derive the CSM density (thus the mass-loss rate) under this ongoing shock interaction model, through the peak luminosity (interaction power) and the peak duration (diffusion time scale) \citep{moriya2014}
\footnote{The required CSM mass is much smaller than the cooling-envelope/CSM case,
as the ongoing shock is much more efficient in the energy conversion without substantial adiabatic cooling \citep[\eg][]{maeda2018}.
Note also that the uncertainty in the luminosity within a factor of two would not alter the mass-loss rate estimate \citep{moriya2014}.}.
The mass-loss rate thus derived is $\sim 4 \times 10^{-3}$~\Msun~\years, assuming the constant wind velocity of $\sim 1,000$~\kms.
Again, the required CSM density (the mass-loss rate) and the size are within the range estimated for the confined CSM around CCSNe.
A drawback in this second scenario (for the low-extinction assumption) is the spectral signature;
the spectrum obtained at $t\sim3$~days reported by \citet{dimitriadis2019} is dominated by a continuum with no trace of narrow emission lines expected for the continuous interaction.
However, the absence of the narrow emission lines does not readily reject the ongoing interaction given a lack of our understanding of the spectral formation in CSM-interacting SNe in general. 

In summary, we are left with two possible interpretations; the cooling envelope/CSM emission (for \ebv$\sim1.0$ mag) or the ongoing CSM interaction (for \ebv$<1.0$ mag). The results of this analysis are summarized in Table~\ref{table:para} together with the properties derived from the second peak,
where $R_{\rm peak}$ is the $R$-band peak magnitude of SN~2019ehk at $t \sim 0$~days.
While the two scenarios involve different mechanisms and result in different natures of the CSM, qualitatively we reach to the same conclusion. There is a dense CSM around SN~2019ehk. This suggests a core-collapse origin for SN~2019ehk. The progenitor might experience (unknown) unstable activity in the final stage of the stellar evolution of the massive star \citep{fuller2018}, but the mass-loss rate derived here might also be consistent with a general expectation from the USSN scenario through the binary interaction (Section \ref{sec:progenitor} for further details). 


\begin{deluxetable*}{lllllll}
\tablewidth{0pt}
\tablecaption{Physical parameters of SN~2019ehk ($M_\ej= 0.43$~\Msun and $\ek= 1.7 \times 10^{50}$~erg irrespective of the extinction). The wind velocity is assumed to be $1,000$ km s$^{-1}$ to estimate the mass-loss rate. In these estimates, $R_v=3.1$ is assumed.}
\tablehead{
  E(B-V) & $R_{{\rm peak}}$ & $L_{{\rm peak}}$ &  $L_{{\rm bump}}$ & M(\Nifs)&  CSM radius & Mass-loss rate\\
 (mag) & (mag) & ($\times 10^{42}$~\ergs) & ($\times 10^{42}$~\ergs) & (\Msun) & (\Rsun) & (\Msun yr$^{-1}$)
}
\startdata
0.5 & -16.29 & 0.583 & 0.783 & 0.022 &11,000 & $4 \times 10^{-3}$\\
0.7 & -16.71 & 0.897 & 1.27 & 0.037 & 9,400  & $4 \times 10^{-3}$\\
1.0 & -17.35 & 1.77  & 2.72 & 0.066 & 1,500  & 1.5 
\enddata
\label{table:para}
\end{deluxetable*}

 \subsection{The USSN as a progenitor system}
 \label{sec:progenitor}

 

The ejecta properties of SN 2019ehk, iPTF14gqr and iPTF16hgs are generally consistent with the expectations for a low-mass ($< 2 M_\odot$) He (or C+O) star explosion corresponding to the main-sequence mass of $8-12$~\Msun, which defines a boundary between the CCSN explosion and a WD formation \citep{kawabata2010}.
In terms of the binary evolution,
such a progenitor can be formed by the USSN scenario with a NS comparison \citep{tauris2013,tauris2015},
which has been actively debated as a promising pathway to form double NS binaries. Indeed, \citet{de2018_14ft} suggested iPTF14gqr as a robust candidate for an USSN. The scenario favored for iPTF16hgs \citep{de2018} is also the USSN, while an eruptive event related to a WD is not rejected. In this paper, we suggest that SN 2019ehk is yet another candidate for an USSN. 

The diversities seen in these objects are in line with the expectation from a core-collapse event of a massive star, as there can be diversities in the progenitor stars, \eg in the progenitor mass and in the properties of the envelope at the time of the explosion \citep{tauris2015}. Indeed, given the low-mass ejecta, a small difference in the progenitor mass in the USSN scenario can easily lead to a large diversity in the observational properties \citep{moriya2017_us,suwa2015,yoshida2017,muller2018}.
For example, \citet{tauris2015} predict that the ejecta mass ranges from
$\sim$ 0.01~\Msun to $\sim$ 0.5~\Msun for the progenitors leading to USSNe.
The range covers the properties derived for SN 2019ehk, iPTF14gqr, and iPTF16hgs.

SN 2019ehk, iPTF14gqr and iPTF16hgs share a unique observational property; they show the `first' peak with a duration of a few days observed in the optical wavelength.
This feature can be naturally interpreted by existence of a substantial amount of CSM,
as has been interpreted for many CCSNe.
The models based on an old system, \eg a WD, have difficulties to create such a short-duration early emission.
The so-called .Ia scenario would create a double-peaked light curve in the `bolometcic' luminosity due to existence of powering radioactive species at different layers. However, in this model, the first peak is generally in the UV and the second peak is in the optical; the optical light curve is indeed predicted not to show the double peaks \citep{shen2010}.
An exception is the double-detonation model of a sub-Chandrasekhar mass WD, which could create the double peaks in the optical \citep{jiang2017,maeda2018}.
However, given the relatively low luminosity of SN 2019ehk and the other objects (as compared to SNe Ia), this will result in the substantial absorption by various Fe-peak lines at the second peak \citep{maeda2018},
and the resulting spectrum would never resemble that of an SN Ib. 

Through the `flash spectroscopy', \citet{de2018_14ft} suggested the existence of a confined CSM around iPTF14gqr. For iPTF16hgs, the properties of the first peak are explained by the cooling-envelope emission from a He-rich envelope without a need of a dense CSM. The first peak observed for SN~2019ehk can be explained either by a very dense and confined CSM (if the extinction is high) or a moderately dense CSM (if the extinction is low, see Section~\ref{sec:bump}). The mass-loss rate derived for the second scenario ($\sim 4 \times 10^{-3}$~\Msun~\years for $\sim 1,000$~\kms) is within the range expected for the mass loss rate associated with the Roche lobe mass transfer in the USSN scenario \citep{tauris2015}. 

Either way, irrespective of an association with a He-rich envelope or a confined CSM, or even the Roche lobe mass transfer, the double-peak light curve suggests a link to the CCSN, especially to the USSN. Given the classification of SN~2019ehk and iPTF16hgs as Ca-rich transients, as well as the similarities of iPTF14gqr to these objects, the existence of multiple populations is indicated within the Ca-rich transients (see Section~\ref{sec:pop}).

\subsection{A subpopulation within Ca-rich transients?}
\label{sec:pop}
 
 \begin{deluxetable*}{llllllll}
\tablewidth{0pt}
\tablecaption{Properties of Ca-rich transients}
\tablehead{
Objects  &  $M$($^{56}$Ni) & Double peak & Environments & Refs \\
         & (\Msun)         &            &              &  }
\startdata
SN 2019ehk & 0.022-0.066 & YES & Young & this paper \\
iPTF16hgs & 0.008       & YES & Young & \citet{de2018} \\
iPTF15eqv & 0.05-0.07    & --- & Young & \citet{milisavljevic2017} \\
iPTF14gqr & 0.05         & YES & Old/Young? & \citet{de2018_14ft} \\
PTF10iuv\footnote{see Section \ref{sec:spec}} &  0.016        & NO  & Old   & \citet{kasliwal2012}
\enddata
\label{table:ca-rich}
\end{deluxetable*}
 

Most of Ca-rich transients show a large offset from a host galaxy \citep{lunnan2017}. This indicates old stellar population as their origin, \eg an explosion or an eruptive event of a WD. \citet{kasliwal2012} and \citet{lunnan2017} showed that the (canonical) Ca-rich transients have homogeneous luminosity distribution, indicating the similar \Nifs masses. The evolution in the spectra and light curve also shows homogeneous natures \citep[Section~\ref{sec:result}; see also][]{de2018}. 

However, there are some Ca-rich transients, including SN 2019ehk, which show a star-forming activity in their explosion sites. SN 2019ehk is on a dust lane along a spiral arm of M100. 
Another Ca-rich transient iPTF16hgs is located at $\sim6$ kpc away from the core of a star-forming galaxy \citep{de2018}, which is atypical as a site of the Ca-rich transient. 
As yet another example, a peculiar Ca-rich transient iPTF15eqv is located on an arm of a spiral galaxy. Early phase data for iPTF15eqv are missing as it was discovered after the maximum, but it is definitely much brighter than the canonical Ca-rich transients with the estimated \Nifs mass of \MNifs$=0.07$~\Msun \citep{milisavljevic2017}. 
Among the USSN candidates discussed in this paper, iPTF14gqr does not clearly show a star-forming activity in its explosion site; it is offset by $\sim 30$ kpc from the core of a star-forming galaxy (or $\sim 15$ kpc from the nearest spiral arm), which is consistent with the old environment generally derived for the Ca-rich transients, while the host galaxy shows a tidally interacting environment \citep{de2018_14ft}.


SN 2019ehk, iPTF16hgs and iPTF14gqr share a distinct feature in their light curves; the first peak with a duration of a few days to a week. This is not seen in the other canonical Ca-rich transients. The analysis of the first peak suggests that they are possibly associated with the USSN scenario. They at the same time show some diversities in the observational properties. iPTF16hgs is nearly indistinguishable from the canonical Ca-rich transients except for the first peak and its potentially young environment.
iPTF14gqr shows quite different characteristics from the Ca-rich transients, 
\ie its early-phase spectra and the double-peaked light curve,
while it evolves to resemble the canonical Ca-rich transients in a late phase.
The almost feature-less spectra of iPTF14gqr around the peak are likely due to
the high luminosity and temperature as well as the high expansion velocities.
The properties of SN 2019ehk are in between iPTF16hgs and iPTF14gqr;
spectroscopically it is indistinguishable from the canonical Ca-rich transients, but it shows a high luminosity, a clear first peak, and a star-forming environment. 

As a consequence of the diversity in the observational properties, the inferred properties of the ejecta and circumstellar environment are also diverse. The ejected \Nifs masses span from \MNifs$=0.01$~\Msun for iPTF16hgs to \MNifs$=0.05$~\Msun for iPTF14gqr.
SN 2019ehk is in between.
The ejecta mass may range from $\sim 0.2 M_\odot$ \citep[iPTF14gqr;][]{de2018_14ft} to $\sim 0.4$~\Msun \citep[iPTF16hgs and SN~2019ehk;][and this paper]{de2018}. There also seems to be a diversity in the nature of the circumstellar environment (Section~\ref{sec:bump}). 

In summary, we suggest that SN2019ehk, iPTF16hgs, and iPTF14gqr (and potentially iPTF15eqv) form a subpopulation within the Ca-rich transient class. They not only explode within the young environment (with a possible exception of iPTF14gqr) but also have observational properties beyond the diversity seen in the canonical Ca-rich transients. Their properties are summarized in Table \ref{table:ca-rich}, together with those of PTF10iuv as a representative of the canonical Ca-rich transients. We identify 4 such objects (SN 2019ehk, iPTF14gqr, iPTF15eqv and iPTF16hgs),
\ie $\sim$10\% of the whole sample of Ca-rich transients.
The rate of Ca-rich transient is $\sim$10\% of CCSNe.
Therefore, the rate of this subpopulation (including SN~2019ehk) is roughly $\sim$1\% of CCSNe.
This rate is consistent with the expectation for the USSN scenario to be a main evolutionary pathway toward double NS binaries \citep{tauris2013,tauris2015}.

\section{Summary} 
 
We performed the optical and near-infrared observations of a luminous Ca-rich transient SN 2019ehk. While it was initially reported as an SN Ib, it shows a rapid development of \caii and Ca~\sii emission lines. The overall spectral evolution matches well to that of the Ca-rich transients. It is thus definitely classified as a Ca-rich transient. Note that normal SNe Ib and Ca-rich transients can be indistinguishable in the early-phase spectra, and thus the change in the classification is not surprising. 

Despite the spectral similarity, SN 2019ehk shows clear differences in its light curve properties from the canonical Ca-rich transients. The peak magnitude is brighter than the typical value seen in the Ca-rich transients by up to 2~mag, depending on the uncertain host extinction.
Furthermore, SN 2019ehk shows a clear first peak which is not observed for the canonical Ca-rich transients. 

From the properties of the second peak, assuming it is powered by the ${}^{56}$Ni/Co/Fe decay chain, we derived the ejecta mass and the explosion energy as $M_\ej= 0.43$~\Msun and $\ek= 1.7 \times 10^{50} $~erg, respectively.
The mass of \Nifs has a large uncertainty due to the host extinction, but is constrained to be $0.022 - 0.066$~\Msun. We interpret the origin of the first peak as the emission associated with a dense, and potentially confined, CSM whose property is within the range inferred for CCSNe. These properties suggest that SN 2019ehk is a variant of CCSNe. Indeed, these properties are largely consistent with the expectations from an explosion of a low-mass He star ($\sim 2$~\Msun), \ie yet another candidate for an USSN. 


We identify at least three (peculiar) Ca-rich transients (SN 2019ehk, iPTF16hgs, and iPTF14gqr) which show peculiar properties beyond the diversity within the canonical Ca-rich transients. As a distinguishing feature, they all show a double-peaked light curve, indicating a CCSN origin. Interestingly, two of them (SN 2019ehk and iPTF16hgs) have the environment atypical as a Ca-rich transient, indicating that their progenitor stars belong to young population. The environment of iPTF14gqr is within the diversity seen for the Ca-rich transients, but its association to the young environment is not rejected \citep{de2018_14ft}. While the pre-maximum data are missing, a peculiar Ca-rich transient iPTF15eqv also shares some properties with these peculiar Ca-rich transients. 

We suggest that these Ca-rich transients form a sub (young) population within the Ca-rich transient class. Their properties can be explained by an explosion of a low-mass ($\sim 2 M_\odot$) He star, i.e., the USSN scenario, and this scenario explains why the observational properties within this subpopulation is much more diverse than those seen in the canonical Ca-rich transients. This suggests an interesting direction in the search of the USSN candidates to understand the evolutionary pathway toward double NS binaries:
high cadence searches for transients to catch the first distinct peak, and continuous followup observations to see the development of the Ca-rich transients' signatures in the late phase. 







\acknowledgments
This research has made use of the NASA/IPAC Extragalactic Database (NED) which is operated by the Jet Propulsion Laboratory, California Institute of Technology, under contract with the National Aeronautics and Space Administration. This research has made use of the NASA/ IPAC Infrared Science Archive, which is operated by the Jet Propulsion Laboratory, California Institute of Technology, under contract with the National Aeronautics and Space Administration.
Keiichi Maeda acknowledges support provided by Japan Society for the Promotion of Science (JSPS) through KAKENHI Grant JP17H02864, JP18H04585 and JP18H05223.
Masayuki Yamanaka is supported by the Grants-in-Aid for Young Scientists of the Japan Society for the Promotion of Science (JP17K14253).
Koji S. Kawabata acknowledges support provided by Japan Society for the Promotion of Science (JSPS) through KAKENHI Grant JP18H03720.
Umut Burgaz acknowledges the support provided by the Turkish Scientific and Technical Research Council (T\"UB\.ITAK$-$2211C and 2214A)
Tomoki Morokuma acknowledges support provided by Japan Society for the Promotion of Science (JSPS) through KAKENHI Grant JP16H02158 and JP18H05223.




\bibliographystyle{apj}
\bibliography{supernova}

\end{document}